\documentclass{camera}

\usepackage{graphicx} 

\begin{document}

\title{In-medium properties of nuclear fragments at the liquid-gas phase 
coexistence} 

\author{A.S.~Botvina} 

\vspace{-10mm}

\organization{Institute for Nuclear Research, Russian Academy of Sciences,
117312 Moscow, Russia}

\maketitle

\vspace{-5mm}

\begin{abstract}
Reactions of nuclear multifragmentation of excited finite nuclei can 
be interpreted 
as manifestation of the nuclear liquid-gas phase transition. 
During this process the matter at subnuclear density clusterizes into 
hot primary fragments, which are located in the vicinity of other 
nuclear species. In recent experiments there were found evidences 
that the symmetry and surface energies of primary fragments change 
considerably as compared to isolated cold or low-excited nuclei. 
The new modified properties of primary fragments should be taken into 
account during their secondary de-excitation. 
\end{abstract}


Multifragmentation has been observed in nearly all types of high
energy nuclear interactions induced by hadrons, photons, and heavy
ions (see a review \cite{SMM}). This is an universal phenomenon
occurring when a large amount of energy is deposited in a nucleus,
and a hot blob of nuclear matter is formed. 
The matter will expand to the sub-saturation densities,
where it becomes unstable and breaks up into many fragments. 
Multifragmentation is a fast process, with a 
characteristic time around 100 fm/c, where, however, 
a high degree of equilibration can be reached. 
For this reason, multifragmentation opens a
unique experimental possibility for investigating the phase diagram of 
nuclear matter at temperatures $T \approx 3-8$ MeV and densities around
$\rho \approx 0.1-0.3 \rho_0$ ($\rho_0 \approx 0.15$ fm$^{-3}$ is
the normal nuclear density). These conditions are typical for the
liquid-gas coexistence region, and they are reached in the 
freeze-out volume at multifragmentation. It is interesting that similar
conditions are realized in stellar matter during the supernova
collapse and explosion \cite{Botvina04,Botvina05}.


\begin{figure}
\centering
\includegraphics[height=7cm]{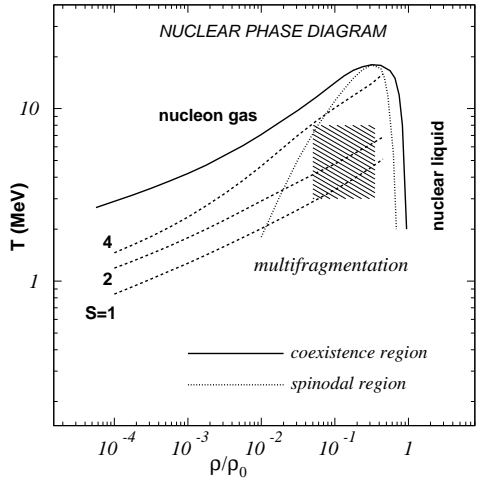}
\caption{Nuclear phase diagram on the temperature--density 
plane. Solid and dotted lines give borders of the liquid-gas 
coexistence region and the spinodal region. The shaded 
area corresponds to conditions reached in nuclear multifragmentation 
reactions. The dashed lines are isentropic trajectories characterized 
by constant entropies per baryon ($S/B=$1, 2, and 4).}
\end{figure}


In Fig.~1 we demonstrate a schematic phase diagram of 
nuclear matter which has a liquid-gas phase transition. The shaded area 
at subnuclear densities 
indicates the region of densities and temperatures which can be studied 
in nuclear multifragmentation processes. We have also shown isentropic 
trajectories with S/B values of 1, 2, and 4 typical for stellar processes 
leading to supernova explosions. 
One can see, for example, that a nearly 
adiabatic collapse of the massive stars with typical entropies 
of 1-2 per baryon passes exactly through the multifragmentation 
area. 



The Statistical Multifragmentation Model (SMM) \cite{SMM} is very 
successful in description of experimental data. 
It is based on the assumption of statistical equilibrium between 
produced fragments in a low-density freeze-out volume. 
We believe that at this point the chemical equilibrium is 
established, i.e., the baryon composition (mass and charge) of 
primary fragments is fixed. However, the hot primary fragments are 
formed in close vicinity to each other, and, therefore, they are
still subject to Coulomb and, possibly, residual nuclear
interactions. Hence, the fragment energies and densities are 
affected by this interaction. 
It is commonly accepted that the liquid-drop
description of individual nuclei is very successful in nuclear
physics. In the SMM the fragments with $A > 4$ are treated as
heated nuclear liquid  drops, and their individual free energies
$F_{AZ}$ are parameterized as a sum of the bulk, surface, Coulomb
and symmetry energy terms:
\begin{equation}
F_{AZ}=F^{B}_{AZ}+F^{S}_{AZ}+E^{C}_{AZ}+E^{sym}_{AZ}.
\end{equation}
In this standard expression $F^{B}_{AZ}=(-W_0-T^2/\epsilon_0)A$ is
the bulk energy term including contribution of internal
excitations controlled by the level-density parameter
$\epsilon_0$, and $W_0 = 16$~MeV is the binding energy of 
nuclear matter.
$F^{S}_{AZ}=B_0A^{2/3}((T^2_c-T^2)/(T^2_c+T^2))^{5/4}$ is the
surface energy term, where $B_0=18$~MeV is the surface coefficient, 
and $T_c=18$~MeV is the critical temperature of infinite
nuclear matter. The Coulomb energy is $E^{C}_{AZ}=cZ^2/A^{1/3}$,
where $c$ is the Coulomb parameter obtained in the Wigner-Seitz
approximation, $c=(3/5)(e^2/r_0)(1-(\rho/\rho_0)^{1/3})$, where
$r_0$=1.17 fm, and the last factor
describes the screening effect due to presence of other fragments.
$E^{sym}_{AZ}=\gamma (A-2Z)^2/A$ is the symmetry energy term,
where $\gamma = 25$~MeV is the symmetry energy coefficient. These
parameters are taken from Bethe-Weizs\"acker formula and
corresponding to the isolated fragments with normal nuclear
density. This assumption has been seen to be quite successful in
description of experimental data concerning charge 
and thermodynamical characteristics of produced fragments 
[4--10]. 
However, in a multi-fragment system in the freeze-out
volume, the parameters of the liquid-drop model may change as
compared with those for isolated nuclei.


\begin{figure}
\centering
\includegraphics[height=6cm]{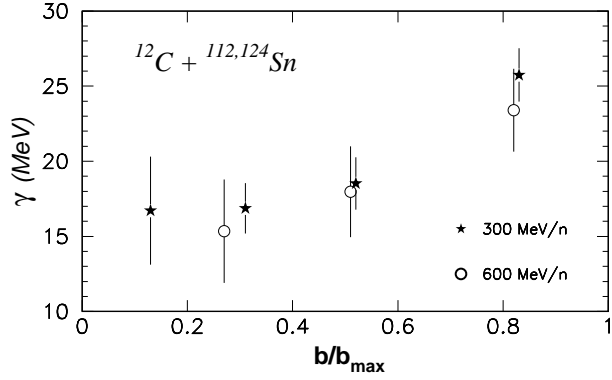}
\caption{The apparent symmetry energy coefficient
$\gamma$ of hot nuclei, as extracted from multifragmentation of tin
isotopes induced by $^{12}C$ beams with energy 300 and 600
MeV per nucleon, versus relative impact parameter $b/b_{max}$
\cite{LeFevre}.}
\end{figure}

\begin{figure}
\centering
\includegraphics[height=7cm]{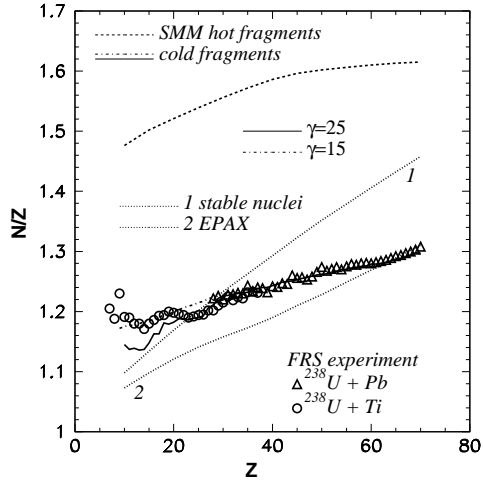}
\caption{
Mean neutron-to-proton ratio versus charge of fragments produced in
multifragmentation-like break-up of $^{238}$U with energy 1 GeV/nucleon
on Pb and Ti targets \cite{Botvina05}. Points are experimental data
obtained on Fragment
Separator (FRS) at GSI. The dashed line is SMM calculation for
primary hot fragments, solid and dot-dashed lines are fragments after
secondary de-excitation. Dotted line 1 corresponds to stable nuclei,
dotted line 2 is the EPAX phenomenological parametrization for nuclei
produced by spallation. The solid and dashed
lines are calculations at the standard symmetry energy parameter
$\gamma = 25$ MeV, the dot-dashed line is for reduced $\gamma = 15$ MeV.}
\end{figure}

In recent years new experiments related to the nuclear isospin and 
isotope distributions in multifragmentation reactions have been performed 
[11--13]. 
The statistical analyses of the isotope data have led to a conclusion 
that modifications of the liquid-drop 
parameters of hot fragments produced in the freeze-out volume are 
needed to explain the experiments. It was suggested that 
the physical reason of this effect could be 
a new physical environment where fragments are formed. The residual 
interactions may lead to energy and density changes which can
effectively be explained by a modification of the macroscopic
nuclear parameters. In particular, it is consistent with predictions 
of many dynamical models that hot fragments formed in a dilute nuclear 
matter have a reduced density too. As known from equations of state, the 
symmetry energy of nuclei decreases at subnuclear densities \cite{Fuchs}. 
The symmetry energy coefficient $\gamma$ was
investigated in the INDRA/ALADIN experiment \cite{LeFevre} by using 
the isoscaling phenomenon. As demonstrated in 
Fig.~2, the coefficient $\gamma$ is about 25 MeV for peripheral 
collisions, as known for cold isolated nuclei. 
It drops down to $\approx 15$ MeV 
for more central collisions leading to multifragmentation reaction. 
The same evolution of $\gamma$ for hot primary fragments was 
also extracted from the TAMU experimental data obtained in central 
and peripheral nucleus-nucleus collisions around Fermi energy 
\cite{Iglio,Souliotis}. In these experiments, besides the isoscaling,  
the neutron-to-proton ($N/Z$) ratio in produced fragments, and the 
fragments isotope distributions were analysed. 
Other high quality experimental data on mean $N/Z$ ratio of 
produced fragments, obtained in a FRS experiment at GSI, are 
demonstrated in Fig.~3. The best description of intermediate 
mass fragments (IMF: $Z=3-20$), which are produced mainly at 
multifragmentation, can be obtained with the SMM at the reduced $\gamma$ 
also \cite{Botvina05}.
It is important that all experimental analyses of the isotope 
characteristics come to the conclusion that 
there is a decreasing of the 
coefficient $\gamma$ to around 15 MeV for hot primary fragments at
multifragmentation conditions. 

A recent analysis of the ALADIN data has revealed modifications in
the nuclear surface properties too \cite{Botvina06}. The 
$N/Z$ dependence of the surface energy was investigated 
for different event classes corresponding to different
excitation energies. At low excitation energies, corresponding to
the onset of multifragmentation, the surface energy follows the
trend predicted by the standard liquid-drop model, i.e., it
decreases with $N/Z$, see Fig.~4. This trend is usually explained by the
contribution of the surface part of the symmetry energy. In the
region of developed multifragmentation, 
where IMF are mostly produced, the surface energy coefficient
becomes nearly independent of the $N/Z$ ratio. Taking into account
this result we conclude that subdivision of the total symmetry
energy into volume and surface parts becomes irrelevant at
multifragmentation conditions. We point out that similar
conclusions about changing surface properties of nuclei were
obtained within the dynamical AMD model \cite{Ono04}. Therefore,
the observed decrease of the symmetry energy of fragments should
not be related to the increase of the total surface of fragments
at multifragmentation. The new surface properties complement 
the medium modification of $\gamma$ coefficient in
the system of many fragments.

At the last stage of the multifragmentation process hot primary
fragments undergo de-excitation and propagate in the mutual Coulomb
field. In the beginning of the de-excitation the hot
fragments are still surrounded by other species, and, therefore,
their modified properties should be taken into account. As far as
we know, only one evaporation code was designed, which takes into
account the modified properties of fragments in their
de-excitation. It was developed in refs. \cite{nihal},
where modifications of symmetry energy were explicitly considered.
Namely, the secondary de-excitation 
starts from the modified symmetry energy of hot nuclei and restore
their normal properties by the end of the evaporation cascade. 
The energy and momentum conservation
laws were fulfilled in the course of this evaporation process.
We emphasize that this de-excitation stage should 
be consistent with the physical mechanism of production of primary 
fragments. For example, a failure to describe the experimental 
isoscaling data \cite{TXLiu04}, by using a dynamical BNV model for 
primary fragment 
formation, may be related to the fact that the sequential evaporation 
of the fragments was included without paying attention to their 
reduced densities, and, as a result, to their small symmetry energies. 
As was demonstrated in \cite{nihal,Iglio} 
this can influence isotope distributions, 
because separation energies of neutrons and protons change essentially. 
We believe that besides the isotope information and isoscaling, 
the relevant analyses should include the corresponding fragment charge 
distributions, IMF multiplicities, temperatures and other information 
[4--10, 15]. 

\begin{figure}
\centering
\includegraphics[height=7cm]{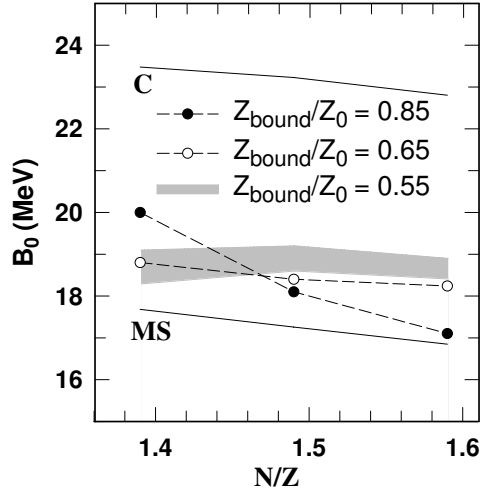}
\caption{
The extracted surface energy coefficient $B_0$ for ensemble sources with
different $N/Z$ ratios at the onset ($Z_{\rm bound}/Z_0$=0.85 bin), and at
the region of full multifragmentation ($Z_{\rm bound}/Z_0$=0.65 and 0.55
bins) \cite{Botvina06}.
The width of the shaded band represents limits given by our
method for the lowest bin.
The $B_0$ obtained from Cameron (C) and Myers-Swiatecki (MS) mass
formulae for cold nuclei are shown by solid lines for
illustration.}
\end{figure}



We conclude that statistical models, which are successfully applied 
for description of nuclear multifragmentation, can be used for 
extended analyses of the hot fragments embedded in surrounding of 
other nuclei and nucleons, typical for the liquid-gas phase 
coexistence. 
Input parameters of the statistical 
models (e.g., the symmetry and surface energies of primary 
fragments) can be directly 
extracted from multifragmentation experiments. There are evidences 
of reduction of the fragment symmetry energy, and modification 
of their surface energy, which are obtained in analyses of 
independent experiments, and they are consistent with model predictions. 
It is important that 
for self-consistent description of the whole process the 
secondary de-excitation of primary fragments should take into 
account the modified properties, at least, in the beginning of 
evaporation cascade. 
One of the goals of future multifragmentation experiments is to 
determine properties of fragments, which are formed in hot nuclear 
systems at subnuclear densities. 

The author thanks I.N.~Mishustin and W.~Trautmann for fruitful 
collaboration, support, and help in preparation of this contribution.

\end{document}